# Modeling Delay Jitter Distribution in Voice over IP

B.Kaushik[a] R.Ganesh [b] and R.Sadhu [c]

[a]Dept of Electrical Engineering, Indian Institute of Technology,
Bombay 400076, India

[b]Dept of Computer Science and Engineering, Indian Institute of Technology,
Bombay 400076, India

[c]Dept of Computer Science and Engineering, Indian Institute of Technology,
Kharagpur 721302, India

It has been suggested in voice over IP that an appropriate choice of the distribution used in modeling the delay jitters, can improve the play-out algorithm. In this paper, we propose a tool using which, one can determine, at a given instance, which distribution model best explains the jitter distribution. This is done using Expectation Maximization, to choose amongst possible distribution models which include, the i.i.d exponential distribution, the gamma distribution etc.

## 1. Introduction

Voice over IP involves various factors and constraints. One of them being the network constraint. This network overhead is manifest in the form of delays in packet arrivals at the destination. Typically, the delay is characterized by a jitter, meaning that the delay characteristics are non-uniform over packet arrivals.

It has been explored in [3] and [1] that if the delay characteristics are captured in an optimal model, then, several interesting characteristics can be determined. One of them being the computation of the optimal play-out time. **[1] discusses how the jitter is given by the variance of the distribution model whereas the first order moment gives the mean delay spacing between packets.** Also [3] discusses how the clock-skew can be computed given a distribution model for the jitter.

[3] discusses that there are several plausible distribution models. Few have been discussed as in [3] and [1]. In this paper, we give an approach for multiplexing between one of these models.

## 2. Notations

Consider the investigated scenario. Two terminals, A and B communicate over a packet-switched network. The actual packet arrival times at B is given by sum of the minimum transmission delay from A to B over the network and $v(t)$ where $v(t)$ is random variable characterizing the extra delay added by the network (delay jitter) at time $t$.

## 3. The Approach

We intend to get a better understanding of the arrival process of audio packets at the receiver. The objective is to ascertain if the packets arriving at a receiver pertain to any well-known distribution, such as exponential, or gamma, or geometric. We note the corresponding distributions below :

[3] models $v(t)$ as an independent, identically distributed (i.i.d) random process with exponential probability density function.

$$f_v(v) = u(v)\mu exp(-\mu v) \qquad (1)$$

where $u(v)$ is the unit step function.

On the other hand, the gamma distribution is given as

$$f_v(v) = \frac{1}{b^a \Gamma(a)} v^{a-1} e^{-\frac{v}{b}} \qquad (2)$$

The idea that the gamma distribution may model the delay jitter distribution, comes by closely comparing the p.d.f for the gamma distribution with the histogram of the jitter history. They match very closely.

In the case of modeling the delay jitters, various authors have suggested different possible models for the delays. The intuition we get from this is that **the network exhibits different behavior at different times and thereby different models for the delay jitters are exhibited at different times**. All that we can observe at the receiver end is therefore only the delay jitters. What remains hidden is *which statistical distribution the jitters correspond to*, at the present(beginning over last n packets say). Thus, we have hidden variables, in the form of indicator variables, which suggest, which section of data over the recent history corresponds to which distribution.

### 3.1. The Expectation Maximization Algorithm

We will now introduce the Expectation Maximization algorithm, better known as the EM algorithm.

In many practical learning settings, only a subset of the relevant instance features might be observable. The EM algorithm is a method to do maximum likelihood estimation in such a setting.

The idea behind EM is

- we have some observed data, but, maximum likelihood estimation of our model is complicated.

- but maybe if there were some extra variables, the maximum likelihood estimation in the "augmented space" would be much simplified.

- the extra variables may be hypothetical.

.Interested readers are referred to [2]. A natural application of EM is for missing data problems.

Therefore, this algorithm provides an effective tool to determine from the given data and a given set of possible distribution functions, which distribution function, each point of the given data actually belongs to.

### 3.2. The algorithm to determine the correct model

Suppose we have a history file showing the delay jitter of a certain number of packets in the recent past, say Count=30,000 (which was typically the number of packets in the trace-files we had for our experiments). Suppose we index these packets with $j$ where $j$ varies from 1 to Count. Let $v(j)$ be the observed delay jitter for the $j^{th}$ packet in that sequence . Let $i$ correspond to the $i^{th}$ model. For the time being, we will focus only on 2 values of $i$, 1 for exponential distribution and 2 for gamma distribution. Let $z_{ij}$ be the indicator variables for the $j^{th}$ packet where $i$ is either 1 or 2.

$$z_{ij} = \begin{cases} = 1 & \text{if the } j^{th} \text{ packet comes} \\ & \text{from the } i^{th} \text{ distribution;} \\ = 0 & \text{otherwise.} \end{cases} \qquad (3)$$

Let

1. $a_i$ = subset of data points from v which come from the $i^{th}$ distribution model

2. $\alpha_i$ = parameter set for the $i^{th}$ distribution model

3. $p_i(v_k | \alpha_i)$ = pdf for data point $v_k$ under the $i^{th}$ distribution model

There are 2 steps for modeling the data

**(I) The E(Expectation) Step** Calculate the expected value of $z_{ij}$ as $Z_{ij}$

$$Z_{ij} = \frac{p_i(v_{ij} | \alpha_i)}{\Sigma_i p_i(v_{ij} | \alpha_i)} \qquad (4)$$

The new values of $z_{ij}$ are calculated as

$$z_{ij} = 1 \ if \ Z_{ij} \geq Z_{ij} \ \forall \ k \ \neq \ i \qquad (5)$$

Then estimate $a_i$ 's as

- $a_i$ = subset of data points v(j) from v for which $z_{ij} = 1$

**(II) The M(Maximization) Step** Find the new Maximum Likelihood Estimates(MLEs) for all the parameter sets $\alpha_i$ using standard methods.

Repeat steps **(I)** and **(II)** above till the $z_{ij}$ values stabilize. Figure 1 shows the pseudocode for the proposed algortihm.

1. /*Read the delay jitter history of N packets*/
   - X = [x(j)] : j = 1 to N
2. /*Suppose the choices for distribution are functions A and B. Initial parameter estimates are [a] for A, [b] for B (obtained by assuming the entire history belongs to one distribution)*/
3. /* z1(j), z2(j) : indicator variables for A, B respectively*/
4. EM algorithm for K iterations till indicator variable stabilizes
5. For i= 1 to K {
   (a) **E step:**
   (b) For j= 1 to N{
       i. z1(j)= p[x(j)—[a]]/{p[x(j)—[a]]+p[x(j)—[b]]}
       ii. z2(j)= 1-z1(j)
       iii. if z1(j) >= z2(j)
           - z1(j)= 1, z2(j)= 0
       iv. else
           - z1(j)= 0, z2(j)= 1
   (c) **M step:**
   (d) for all x(j) where z1(j)=1
       a = parameter estimate of x(j)
   (e) for all x(j) where z2(j)=1
       b = parameter estimate of x(j)
6. end for
7. plot(z1)

Figure 1. The Pseudo-code

## 4. Experiments and results

1. The experiments were performed over trace files used in [1]. There were 6 such trace files. The EM algorithm was run on the data, till the indicator variables stabilized. Below we plot the graph of the first set of indicator variables $z_{1j}$.

2. For the first trace file, the algorithm converged in a single step. The single plot is shown in figure 2

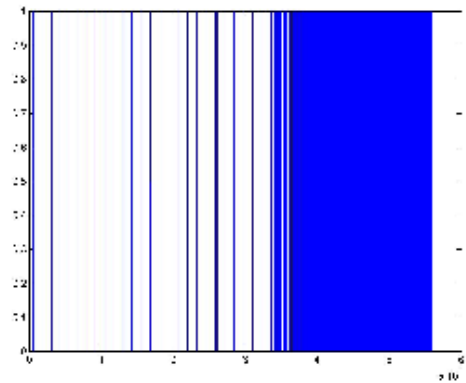

Figure 2. Graph for trace 1 (generated for packets 1-56000)

The figure 2 shows that in the first half, the jitter tends to follow the gamma distribution, whereas, later on it tends to follow the i.i.d exponential distribution.

The algorithm was then run on smaller window of size 3500 which was moved over this trace-file. The corresponding trace-files are shown in figures 3, 4, 5 and 6

The observation is that when we keep a window of size 3500 and move the window around, the transition from gamma to exponential distribution becomes apparent (in fig. 4) around the actual region of change (as in fig. 2), albeit a little delayed. For the regions in which the exponential/gamma distribution predominate, this smaller window is in complete agreement (gamma for fig. 3 and exponential as in fig. 6 ).

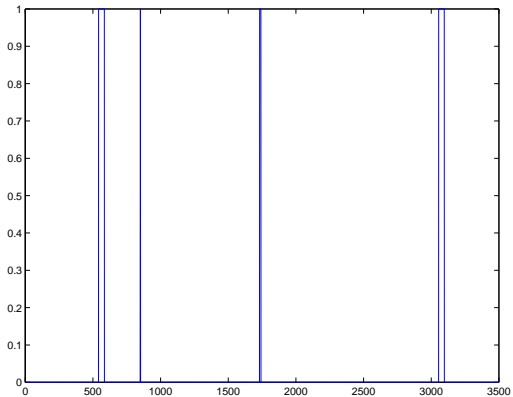

**Figure 3.** Graph for trace 1 (generated for packets 0-3500)

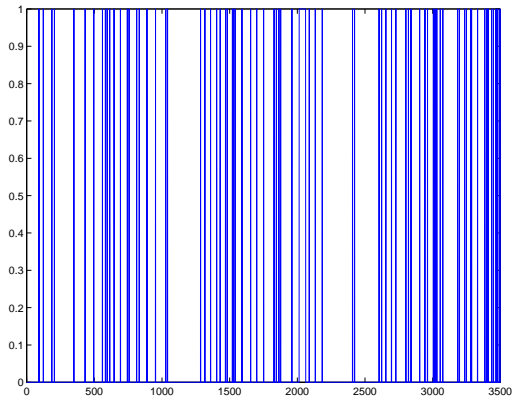

**Figure 4.** Graph for trace 1 (generated for packets 35000-38500)

## 5. Conclusions

In this paper, we have suggested that we can have a way of choosing between alternative models for the delay jitters of packets in voice over IP. It was found that the jitters indeed follow one particular model at a stretch, provided that we have the right models which to choose from. How the variance and mean of the captured model can be used in play out time estimation has already been partly dealt with in [3].

We recommend a middle-tier architecture as shown in figure 7 for capturing the jitter distribution characteristics. The proxy, for instance, can monitor over intervals of say 30,000 packets (or any duration for which the network characteristics are expected not to change). The proxy can determine after periodic intervals, over a window of size, say 3500 packets, the nature of the distribution of the delay jitters. The distribution type and the corresponding parameters can be indicated to the receivers through a special field in the header. This has 3 advantages : *one*, that the individual receivers are not overloaded and *two*, that the proxy can be **continuously monitoring** the network, since it receives voice packets from for different destinations from different sources all the time. *Thirdly*, this architecture greatly reduces the redundancy that is otherwise incurred at each of the receivers trying to estimate the distribution. *Finally*, this also provides a neat way of dynamically updating the receivers on the status of the network at any time.

The question remains, *in which layer should this task of determining the appropriate model, be performed ?*. A possible solution is to do this at the receiver end, by the VOIP software. The software continuously monitors the packets arriving at the receiver end. Periodically, (say, once every 10 seconds- or any time duration, over which the network characteristics generally change), it takes a history of approximately 30,000 packets (if there is no conversation, it can send dummy test packets over the network and monitor their delay time) and over them, it computes the probable model for the present. In this way, given that a call is received at any instance, the system can suggest, based on the history, what model of jitter distribution, the network imposes on the current

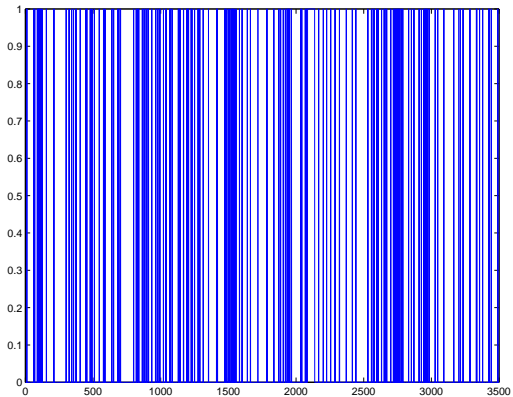

Figure 5. Graph for trace 1 (generated for packets 39000-42500)

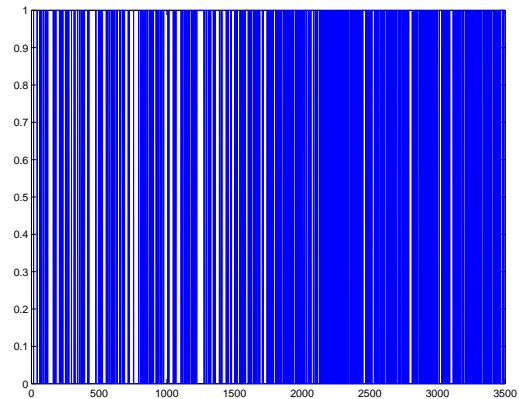

Figure 6. Graph for trace 1 (generated for packets 42500-46000)

conversation.

## REFERENCES

1. Sue B.Moon, Jim Kurose, and Don Towsley, "Packet Audio Play-out Delay Adjustment: Performance Bounds and Algorithms, *Multimedia Systems*, Vol-6, pp.17-28, January 1998.

2. Tom Mitchell, "Machine Learning", *McGraw Hill*, 1997,

3. Tonu Trump, "Estimation of clock skew in telephony over packet switched networks", *Proc. IEEE International Conference on Acoustics, Speech and Signal Processing* , Istanbul, vol.1 pages 2605 to 2608, 2000 pp. 191-196

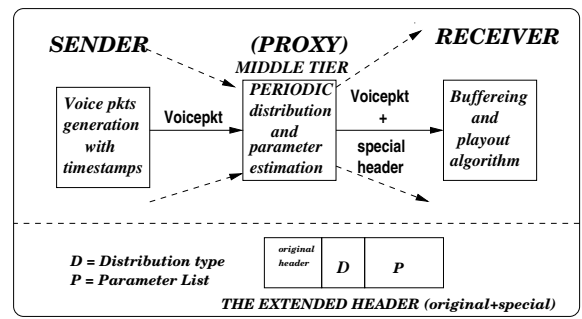

Figure 7. The proposed architecture

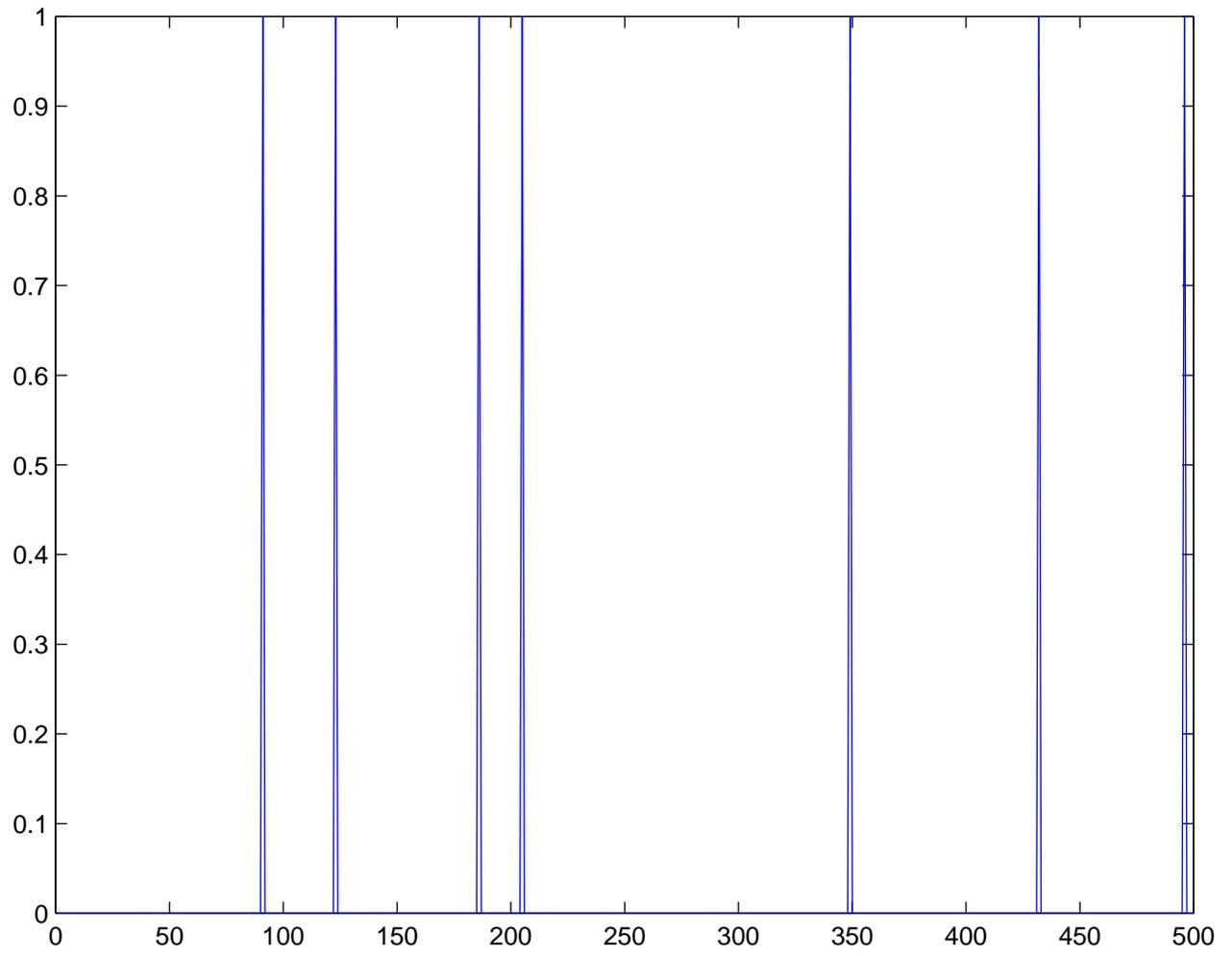

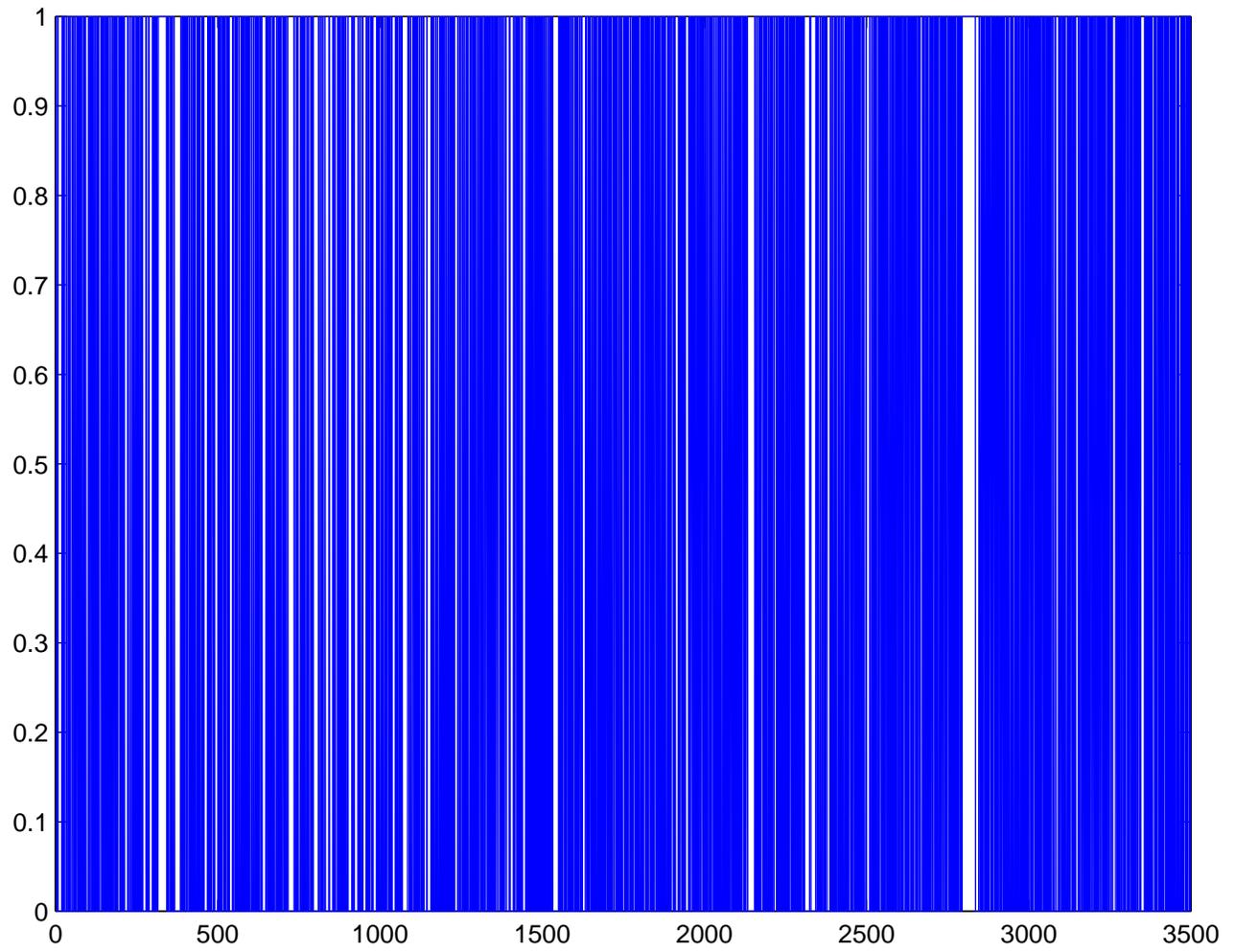

1. **/\*Read the delay jitter history of N packets\*/**

   - X = [x(j)] : j = 1 to N

2. **/\*Suppose the choices for distribution are functions A and B. Initial parameter estimates are [a] for A, [b] for B (obtained by assuming the entire history belongs to one distribution)\*/**

3. **/\* z1(j), z2(j) : indicator variables for A, B respectively\*/**

4. **EM algorithm for K iterations till indicator variable stabilizes**

5. For i= 1 to K {

   (a) **E step:**

   (b) For j= 1 to N{

       i. z1(j)= p[x(j)—[a]]/{p[x(j)—[a]]+p[x(j)—[b]]}

       ii. z2(j)= 1-z1(j)

       iii. if z1(j) >= z2(j)

          - z1(j)= 1, z2(j)= 0

       iv. else

          - z1(j)= 0, z2(j)= 1

   (c) **M step:**

   (d) for all x(j) where z1(j)=1

       a = parameter estimate of x(j)

   (e) for all x(j) where z2(j)=1

       b = parameter estimate of x(j)

6. end for

7. plot(z1)